\documentclass[aps,prb,twocolumn,showpacs,superscriptaddress]{revtex4-1}

\usepackage{graphicx}
\usepackage{dcolumn}
\usepackage{bm}
\usepackage{subscript}
\usepackage{subfig}
\usepackage{amsmath}
\usepackage{ulem}
\usepackage{color}

\newcommand{\hamilton}{\mathcal{H}}
\newcommand{\matrixX}[1]{\bm{#1}}
\newcommand{\eps}{\varepsilon}
\newcommand{\up}{\uparrow}
\newcommand{\down}{\downarrow}
\newcommand{\tesub}[1]{\textsubscript{#1}}

\newcommand{\op}[1]{\operatorname{#1}}

\newcommand{\creOp}[2]{#1^{\dagger}_{#2}} 
\newcommand{\annOp}[2]{#1^{}_{#2}} 

\begin{document}

\title{Dynamically generated edge states in topological Kondo insulators}

\author{Jan Werner}
\email{jwerner@physik.uni-wuerzburg.de}
\author{Fakher F. Assaad}
\affiliation{Institut f\"ur Theoretische Physik und Astrophysik, Universit\"at W\"urzburg, Am Hubland, D-97074 W\"urzburg, Germany}

\date{\today}

\begin{abstract}
Kondo insulators combine strong electronic correlations with spin orbit coupling and thereby provide a potential realization of correlated topological 
insulators.
We present model calculations which allow us to study the onset of bulk coherence and concomitant topological edge states from the mixed valence
to local moment regimes.
Our real-space dynamical mean-field results include the detailed temperature dependence of the single particle spectral function
on slab geometries as well as the temperature dependence of the topological invariant. 
The relevance of our calculations for candidate materials like SmB\tesub{6} is discussed.
\end{abstract}


\maketitle 
The essence of heavy fermion systems is captured by the hybridization of narrow -- nearly localized -- and wide -- delocalized --
electronic bands \cite{Fisk_rev,Coleman_HF_chapter,Tsunetsugu97_rev}. In the {\it high} temperature limit, the two systems
essentially decouple, while at temperatures below the coherence scale a Fermi liquid  emerges. The effective mass of the dressed
quasiparticles, or inverse coherence temperature, can exceed by orders of magnitude the bare electronic mass.
The Kondo insulating state occurs when there is precisely one localized and one delocalized electron per unit cell. It is the
ultimate sign of coherence in the sense that the dynamically induced hybridized band structure is precisely half-filled, with the
chemical potential right inside the hybridization gap.
Based on mean-field theories which capture very well the ground state properties it has been argued \cite{Coleman_PRL_TKI, Coleman_PRB_TKI}
that the Kondo insulating state can be a quantum spin Hall insulator \cite{Hasan10_rev,Zhang_Topo_Ins_RMP}, thus providing an
explicit example of a correlation-induced topological state \cite{Hohenad_Rev_Topo_Insulators}: the topological Kondo
insulating state (TKI).

The origin of the TKI can be traced back to the odd-parity wave-function of the nearly localized $f$-electrons
hybridizing with even parity delocalized conduction electrons alongside strong spin-orbit coupling. 
Although the ground state of the TKI is adiabatically connected to the non-interacting quantum spin Hall state,
correlation effects show up in a non-trivial temperature dependence of spectral functions.
The TKI state provides an interpretation of the low temperature resistivity anomalies observed in many Kondo insulators \cite{Riseborough00}
such as SmB$_6$  \cite{Allen79}, YbB$_{12}$ \cite{Iga98,Batkova06} or even  Ce$_3$Bi$_4$Pt$_3$ \cite{Hundley90}.
Recent ab-initio band structure calculations have classified SmB$_6$ \cite{Lu13} and YbB$_{12}$ \cite{Weng13} as
topological insulator and respectively topological crystalline insulator.
For SmB$_6$ it is experimentally \cite{SmB6_Topo_Exp,SmB6_Spectroscopy_Exp,SmB6_Surface_Hall_Exp,Li13}
concluded that the resistivity saturation originates from surface as opposed to bulk conductance.
Additional evidence comes from the observation of a residual density of states at the Fermi level as the
temperature $T\to0$ \cite{SmB6_gap_STS} as well as from photoemission studies
\cite{Frantzeskakis13,Neupane13,SmB6_charge_fluctuations} which lead to the interpretation of a
Fermi surface consisting of three pockets as appropriate for a topological insulator \cite{Fu07a}.
The aim of this article is to provide a detailed numerical calculation of a minimal model which captures
generic features of the TKI state. 
We will scan from mixed valence to local moment regimes, map out the relevant energy scales, the topological invariant,
and provide a temperature dependence analysis of the single particle spectral function on a slab geometry.
In the   next  two sections we describe   our model and  methods.  Details of the technical implementation  can be found in the Appendix. 
Section \ref{Results}  is devoted to a discussion of our results,  both for bulk and boundary properties  and in Sec. \ref{Conclusions} we  conclude. 
\section{Model.} 
\label{Model}
Following  Ref.~\cite{Coleman_PRL_TKI}  we will  consider a Kramers doublet of almost localized $f$-electrons hybridizing 
with a delocalized conduction band.
The corresponding Hamiltonian \cite{TKI_Transitions_DMFT} is $\hamilton = \hamilton_0 + \hamilton_U$ where
\begin{equation}
  \label{eq_hamiltonian}
  \hamilton_0 = \sum \limits_{k\in BZ}
  \begin{pmatrix} d^{\dagger}_{k} \\ f^{\dagger}_{k} \end{pmatrix}^{T}
  \begin{pmatrix}
     \matrixX{E_d}(k) &  V^{\ast} \matrixX{\Phi}^{\dagger}(k) \\
    V \matrixX{\Phi}(k) & \matrixX{E}_f(k)
  \end{pmatrix}	
  \begin{pmatrix} d_{k} \\  f_{k} \end{pmatrix}
\end{equation}
and $\hamilton_U = U \sum_i n^{(f)}_{i\up} n^{(f)}_{i\down}$.
The operator $d_k = (d_{k,\up},d_{k,\down})$ creates a conduction electron with momentum $k$, while $f_k$ creates
an electron in the almost flat $f$-band.  We consider a two-dimensional (2D) square lattice with 
$\matrixX{E}_d(k) = -2 t_d (\cos(k_x) + \cos(k_y)) \matrixX{1} $   and
$\matrixX{E}_f(k) = \Bigl ( \eps_f-2 t_f (\cos(k_x) + \cos(k_y)) \Bigr ) \matrixX{1} $.
The form factor $\matrixX{\Phi(k)}$ encodes the spin-orbit coupling, which leads
to the non-trivial band topology \cite{Coleman_PRB_TKI}. It reads
$\matrixX{\Phi}(k) = \vec{d}(k) \circ \vec{\matrixX{\sigma}}$,
where $\vec{d}(k) = ( 2 \sin(k_x), 2 \sin(k_y),0 )$ \cite{Tran_PRB_TKI}.
Although spin is not conserved, the Hamiltonian is block diagonal, so that due to time-reversal symmetry 
one can associate a conserved pseudo spin with the two Kramers conjugate blocks.
$\hamilton_U$ describes a Hubbard interaction which acts only on the $f$-orbitals.
Throughout this article we consider $V=0.4 t$, $\eps_{f}=-6.0t$ and $t_d=t$ as well
as a small hopping $t_f=-0.2 t$ compatible with the almost localized character of the
f-orbitals. The negative sign of $t_f$ is necessary for the band inversion,
which induces the topological state \cite{3D_TKI}.
We vary the interaction strength to tune the system to both the mixed valence and local moment regimes.
As appropriate for the Kondo insulating state, we consider the half-filled band case.
In a previous study, we investigated the ground state bulk phase diagram  \cite{TKI_Transitions_DMFT},  thus establishing the  
occurrence of distinct crystalline topological insulating states \cite{Slager2013}.
\par
Here emphasis is placed on dynamical and topological finite temperature properties.
We use periodic boundary conditions to infer bulk properties. In addition to that, we
use a ribbon geometry of finite width $N_y$ with open boundaries in the y-direction,
but periodic boundary conditions in the x-direction, to directly investigate the
dynamically induced topological surface states. 

\section{Numerical methods} 
 The method of choice to study paramagnetic phases of heavy fermion systems is
the Dynamical Mean-Field Theory (DMFT) \cite{RevDMFT96}. The central approximation is the assumption of
a local self-energy. This neglects non-local spacial correlations, while the frequency dependence
is fully retained, and allows to capture the salient features of the paramagnetic heavy fermion
state \cite{PAM_DMFT_Jarrell,Coherence_Metamag_Kondo,Martin10}. By allowing for an inhomogeneous,
site-dependent self-energy, we account for the inequivalence of sites in the open boundary case \cite{PotthoffSDDMFT}.
Thus the lattice problem is mapped to a series of $N_y$ impurity problems, which are coupled
via the DMFT self-consistency condition
\begin{equation*}
\bigl [ \mathcal{G}(i \omega) \bigr ]_{i i} = \bigl [ G_{loc}(i \omega) \bigr ]_{i i},
\end{equation*}
where $\mathcal{G}$ and $G_{loc}$ are the $(4 N_y \times 4 N_y)$-dimensional impurity
and local lattice Green's function, respectively.
The numerically exact CT-HYB \cite{PWernerPRL,PWernerPRB} formulation of the
quantum Monte Carlo algorithm, which becomes advantageous
especially in the strong-coupling regime \cite{PerfAnPRB} is used to solve the resulting, auxiliary impurity problems.
To obtain the self-energy from the impurity calculations, we employ an improved estimator based on higher-order
correlation functions \cite{improved_sigma_Hafermann}, which significantly reduces the
statistical noise. We continue the converged result for the self-energy
to the real-frequency axis \cite{AF_Mott_self_energy,spectral_properties_3D_Hubbard,Goth_magnetic_impurities_KM}
by means of the stochastic analytical continuation method \cite{KBeachMaxEnt}, to obtain the
orbital and momentum resolved spectral function.
\begin{align}
    A_i(k,\omega) &= -\pi^{-1}\, \op{Im} \Bigl [ G(k, \omega^{+})\Bigr ]_{ii}\\
    G(k,\omega^{+}) &= \Bigl ( \bigl (G^{(0)}(k,\omega^{+}) \bigr )^{-1} - \Sigma(\omega) \Bigr)^{-1} \notag
\end{align}
where $\omega^{+} = \omega+i 0^{+}$. Details of our approach can be found in  Appendix \ref{appendix}.\\

\section{Numerical results}
\label{Results}
In this section, we describe in some details our numerical results. The aim is to provide a systematic study of the evolution of relevant 
energy scales from the mixed valence to local moment regimes. 
\subsection{Topological  invariant and coherence.}
\begin{figure}[h!tbp]
    \includegraphics[width=0.49\textwidth]{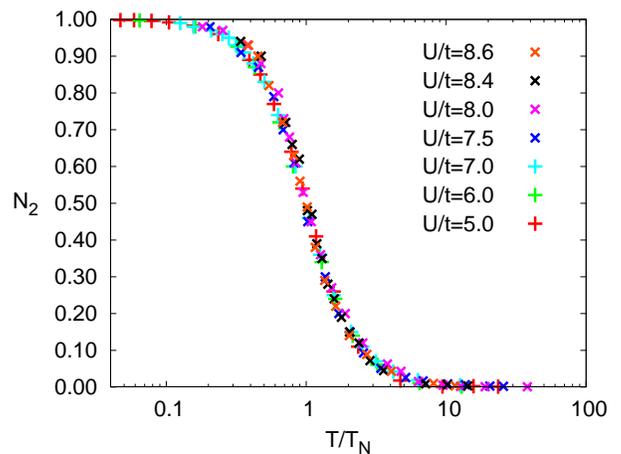}
    \caption{Topological invariant $N_2$ over temperature in units of $T_N$, for different values of $U/t$.}
    \label{fig_N2}
\end{figure}

\begin{figure}[h!tbp]
    \includegraphics[width=0.49\textwidth]{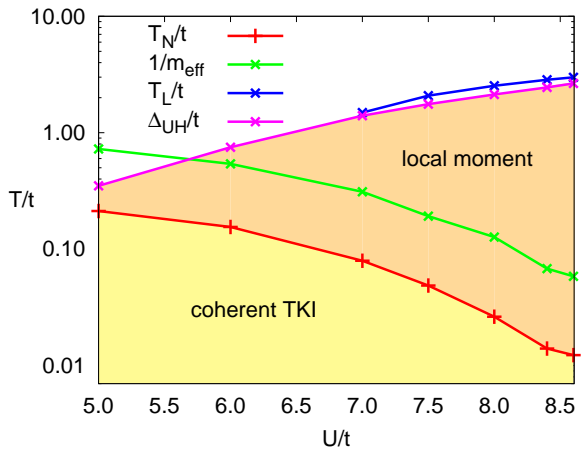}
    \caption{Evolution of the different temperature scales with $U$.}
    \label{fig_scales}
\end{figure}

In TKI states the notion of coherence and the emergence of a
topological band structure are intimately related. Following Ref.~\cite{Corr_Topo_Insulator_BHZ} we have computed
the topological invariant $N_2$ \cite{Volovik_Universe_in_a_He_droplet} which in the framework of the DMFT takes the form: 
\begin{align}
\label{eq_N2}
N_2 = \op{Im} \Bigl [ \frac{d}{d\, \omega} K(\omega^{+}) \Bigr ]_{\omega = 0}  \; {\rm with } \;  
K(i \nu) =   -\frac{1}{V \beta} \sum \limits_{k,i \omega,\sigma}  \\ \frac{\sigma}{2} \op{Tr} \Bigl [
\frac{\partial\, h_{\sigma}(k)}{\partial\, k_y} \notag 
 \, G_{\sigma}(k,i \omega + i \nu)\,
\frac{\partial\, h_{\sigma}(k)}{\partial\, k_x} \, G_{\sigma}(k,i \omega)
\Bigr]\notag
\end{align}
where $\sigma$ is the pseudo spin. Instead of taking the derivative at real frequency $\omega=0$,
we approximate $N_2$ by the finite difference between the zeroth and first Matsubara frequencies.
In the absence of interactions and at $T=0$, $N_2$ corresponds to the quantized pseudo spin Hall
conductivity in units of $e^2/h$. Since the TKI is adiabatically connected to a non-interacting state,
the quantized nature of $N_2$ remains robust in the presence of interactions. 
In  Fig.~\ref{fig_N2} we plot $N_2$ for various values of $U/t$ as a function of temperature in units
of $T_N$. The latter quantity is defined such that 
\begin{align} 
N_2(T=T_N) = \frac{1}{2}
\end{align}
The value of $T_N$ as a function of $U/t$ is given in Fig.~\ref{fig_scales}. As apparent, by plotting $N_2$ in units of $T_N$
we achieve a nice data collapse, thus suggesting that the formation of the topological band structure follows a single scale.
That $T_N$ is indeed nothing but the coherence scale $T_{coh}$ can be seen by computing the bulk 
effective mass defined as
\begin{equation*}
    m_{eff} = 1-\frac{d\, \Sigma_{bulk}(\omega)}{d\, \omega} \Bigl |_{\omega=0} \sim 1/T_{coh}.
\end{equation*}
The comparison of the energy scales in Fig. \ref{fig_scales} reveals that $T_N$ is proportional to the inverse of $m_{eff}$.
A data collapse like the one in Fig.~\ref{fig_N2} suggests to relate $T_N$ to the single impurity Kondo temperature $T_K$. However,
for a lattice model like the PAM, in particular for mixed valence, $T_K$ is not well-defined. Therefore we
avoid the notion of the Kondo scale $T_K$. At least in the local-moment limit, namely in the Kondo lattice model,
numerical studies find that at fixed particle number $T_K$ and the coherence scale are closely related \cite{Coherence_scale_Kondo_lattice,Coherence_vs_Kondo_scale_Assaad}.\\
\subsection{Local moment and mixed valence regimes.}  
 In units of the hopping matrix element $t$ and
for the considered values of the Hubbard interaction the coherence scale varies roughly by an order of magnitude
and interpolates between the mixed valence and local moment regimes. We can confirm this statement by considering the quantity 
\begin{equation}
\label{eq_theta}
\Theta = 1 - \frac{ \bigl < n^{(f)}_{\up} n^{(f)}_{\down} \bigr >}{ \bigl < n^{(f)}_{\up} \bigr > \bigl < n^{(f)}_{\down} \bigr >}
\end{equation}
which is zero in the uncorrelated case, and approaches one for a local moment, since in this limit
$\bigl < n_{\up} n_{\down} \bigr > \to 0$.
\begin{figure}[h!tpb]
   \includegraphics[width=0.49\textwidth]{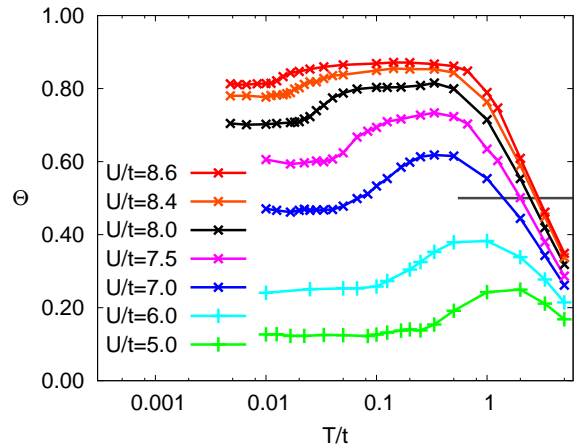}
    \caption{Formation of the local moment as defined by $\Theta$ (see Eq.~\ref{eq_theta}) at  different values of $U/t$. 
       The gray horizontal line indicates the threshold value for the local moment scale.}
    \label{fig_theta}
\end{figure}
As can be seen in Fig.~\ref{fig_theta}, $\Theta$ increases with lowering
of the temperature, and saturates at the lowest temperatures. For sufficiently large $U/t$, there is a plateau 
beyond which $\Theta$ drops slightly. We attribute this drop to the enhanced itineracy of the
$f$-electrons -- translating into a growth of the double occupancy -- arising as coherence sets in.
In fact, the energy scale at which $\Theta$ converges to its zero temperature value matches
approximately the coherence scale.  We define the temperature scale for the crossover
to the local moment $T_L$ by
\begin{equation}
 \Theta(T=T_L) = 1/2.
\end{equation}
 This allows us to obtain values for $T_L$ in the range $U/t \geq 7.0$.  
The so obtained local moment scale, $T_L$, is included in Fig.~\ref{fig_scales}. It turns out
that $T_L$ can be well understood as the energy gap
between the chemical potential and the upper Hubbard band $\Delta_{UH} = \eps_f+U-\mu$,
which is included for comparison. Because of this, we will use $\Delta_{UH}$ as a
measure of the crossover scale to the local moment regime.

\subsection{Temperature dependence of the spectral function.}
\begin{figure*}[htbp]
\begin{center}
  \includegraphics[width=0.9\textwidth]{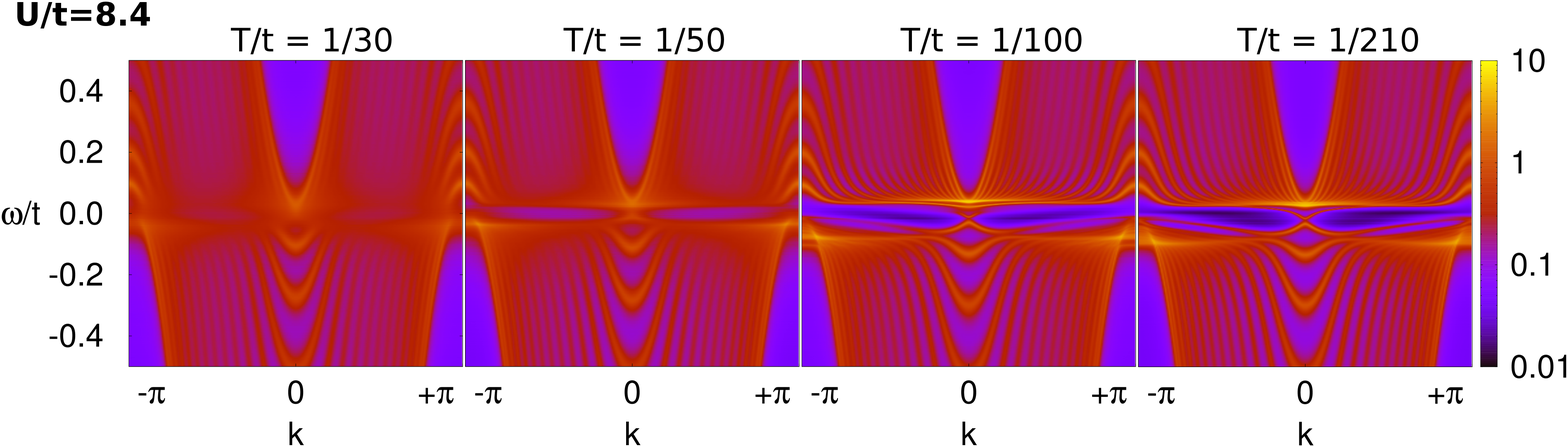}
\includegraphics[width=0.9\textwidth]{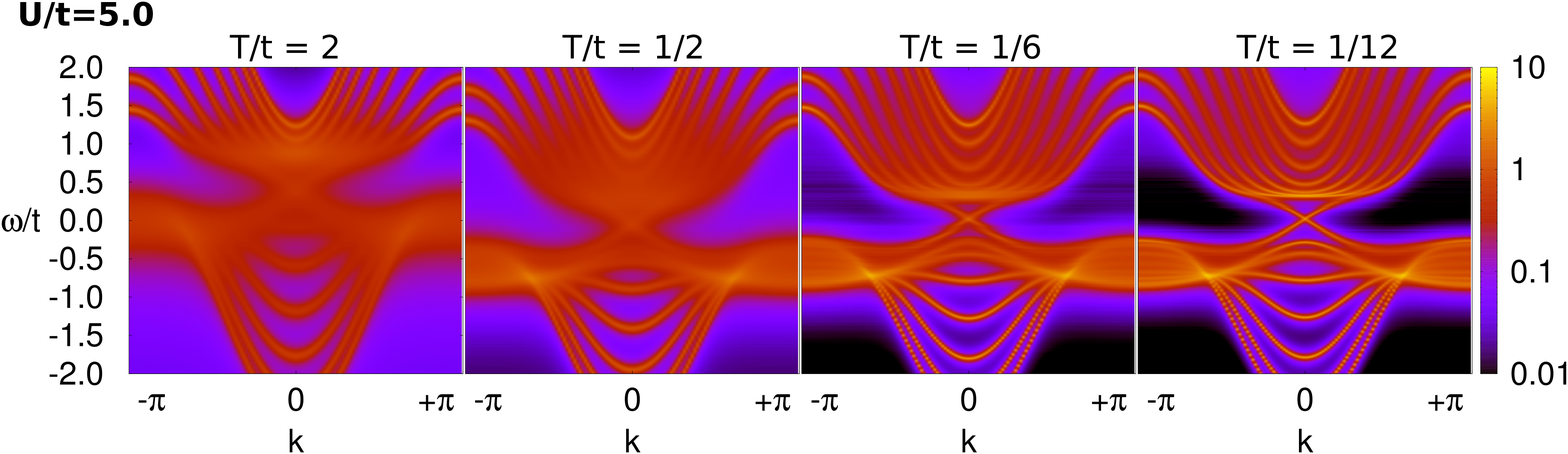}
\end{center}
\caption{Temperature dependence of the single particle spectral function in the local moment
($N_y=16$, upper panel) and mixed valence regimes ($N_y=8$, lower panel),
projected to the one-dimensional surface Brillouin zone. Above the coherence temperature
$T_N/t \approx 0.014$ at $U/t = 8.4 $ and $T_N/t \approx 0.21$ at $U/t = 5$ one observes broad,
incoherent, features out of which the bulk gap and helical surface state emerge as $T$
drops below $T_N$.}
    \label{fig_aom}
\end{figure*}
\begin{figure}[h!tbp]
    \includegraphics[width=0.49\textwidth]{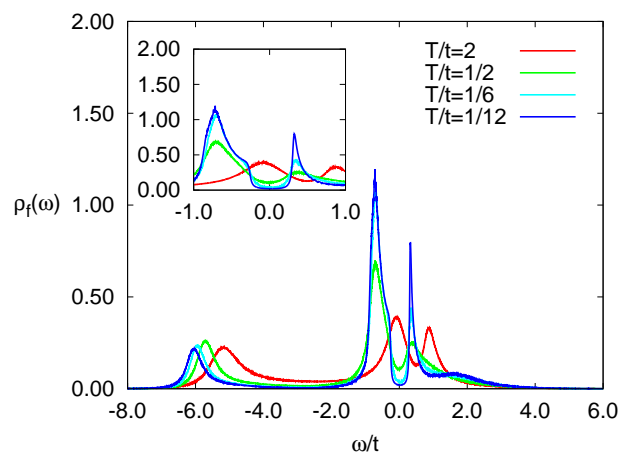}
    \caption{Bulk density of states of the $f$-band as a function of temperature for $U/t=5.0$.}
    \label{fig_dos_U50}
\end{figure}
\begin{figure}[h!tbp]
    \includegraphics[width=0.49\textwidth]{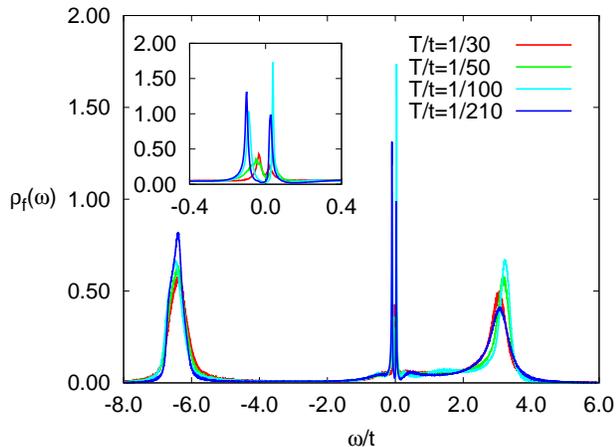}
    \caption{Bulk density of states of the $f$-band as a function of temperature for $U/t=8.4$.}
    \label{fig_dos_U84}
\end{figure}

Fig.~\ref{fig_aom} plots the temperature dependence of the total spectral function, $\sum_{i} A_{i}(k,\omega)$, on slab  
geometries in the local moment ($U/t = 8.4$) and mixed valence ($U/t=5$) regimes. For both considered parameter sets, 
the bulk $f$-density of states is plotted in Figs.~\ref{fig_dos_U50}  and \ref{fig_dos_U84}.  In the local moment regime, 
the lower (upper) Hubbard bands are located well below (above) the Fermi  energy. In contrast, in the mixed valence regime, 
the upper Hubbard  band lies in the  vicinity of the Fermi energy.  The canonical particle-hole transformation, 
$d_{k} \rightarrow d^{\dagger}_{k+Q}$ and $f_{k} \rightarrow f^{\dagger}_{k+Q}$  where $Q = (\pi,\pi)$ maps the
present results to the case where the  
lower Hubbard band lies in the vicinity of the chemical potential and the upper Hubbard band well above
the Fermi energy. In both the mixed valence and local moment regimes, the coherence scale, $T_N$, marks the 
onset of the formation of the bulk band gap and the emergence of the helical edge state. In the mixed valence 
regime, there is a substantial reordering of spectral weight in the vicinity of the Fermi energy and at a temperature 
scale set by the coherence scale. In the local moment regime, coherence is accompanied by a {\it small} transfer of 
spectral weight from the upper and lower Hubbard bands to a region in the vicinity of the Fermi energy.
This transfer of spectral weight reflects the enhanced itinerancy of the $f$-electrons  below the coherence temperature.  
Above $T_N$ incoherent features are detectable in the spectral function. In the local moment regime, this
stems from spin flip scattering of conduction electrons off the well-formed local moments. In the mixed valence regime,
incoherence should be assigned to Mott Hubbard physics \cite{Imada_rev}. Below $T_N$ one observes the Dirac cone
with crossing point in the close vicinity of the Fermi energy. Hence, although the effective mass of the 
f-electrons varies by an order of magnitude, the effective mass of the edge state remains small.
This observation is in accord with recent quantum oscillation experiments of Ref.~\cite{Li13}.
Despite the very different pictures involved in the understanding of the mixed valence and local moment regimes,
the temperature dependence of the low-lying features of the single particle spectral function show
remarkable similarities when the energy is measured in units of the coherence scale 
$T_N$. This is consistent with Fig.~\ref{fig_N2} which demonstrates that $T_N$ is the only energy scale.   


\section{Conclusions}
\label{Conclusions}
In summary, we have studied a model for topological Kondo insulators within real space DMFT+CTQMC.
We have mapped out the relevant scales from the mixed valence to local moment regimes,
and showed that the characteristic temperature scale below which the topological invariant
$N_2$ saturates tracks the coherence temperature.  
Of particular importance for ARPES \cite{Frantzeskakis13,Neupane13,SmB6_charge_fluctuations} and quantum oscillation
\cite{Li13} experiments is the single particle spectral function. We have computed this quantity
on ribbon geometries so as to allow for the emergence of edge states.
Throughout the considered  parameter range, which covers an order of magnitude variation
in the f-electron effective mass, we observe dynamically induced edge states below the bulk coherence temperature.
To a first approximation the Fermi velocity of  the edge state tracks the
coherence temperature (inverse f-electron effective mass) but has a small
effective  mass since the Dirac cone crossing lies in the vicinity of the Fermi  energy.
This large difference between the surface and bulk f-electron effective masses is consistent with observations of  Ref.~\cite{Li13}. 
%
%
%
%
Our model captures the minimal ingredients for the description of the TKI state.
Many of the results -- based only on topological considerations and the purely local Kondo effect -- will
carry over to more realistic situations found in possible candidate materials.  
Qualitative understanding of SmB$_6$ clearly requires to take a step beyond the Kramers doublet description of the 
f-state so as to take into account the cubic symmetry inherent to this material \cite{Alexandrov13} as well as the
chemical reconstruction of the SmB$_6$ surface \cite{Zhu13_SmB6}.

\appendix
\section{ Real-space DMFT}
\label{appendix}
To study the system with open boundaries, we employ the real-space formulation of the Dynamical Mean Field Theory \cite{PotthoffSDDMFT}.
In this approach, the inhomogeneous lattice system is mapped to a series of single impurity problems $j=1\dots N_y$, which can be described by the action
\begin{equation}
\label{action}
S_j = \int \limits_{0}^{\beta} d \tau \, d \tau' \, \sum \limits_{\sigma} \creOp{f}{\sigma}(\tau) \Delta_j(\tau-\tau')\annOp{f}{\sigma}(\tau') + \int \limits_{0}^{\beta} d\tau\, \hamilton_{imp}(\tau).
\end{equation}
The individual impurity problems differ in the hybridization function $\Delta_j$, which is obtained by the self-consistency condition
\begin{eqnarray*}
&  & \Bigl ( i \omega + \mu - \eps_f - \Delta_j(i \omega) - \Sigma_j(i \omega) \Bigr )^{-1} = \bigl [ \mathcal{G}(i \omega) \bigr ]_{j j}  \\ 
& & = \bigl [ G_{loc}(i \omega) \bigr ]_{j j}.
\end{eqnarray*}
$\mathcal{G}$ is the impurity Green's function, while the local lattice Green's function is given by
\begin{equation*}
\matrixX{G}_{loc}(i \omega) = \frac{1}{N_x} \sum \limits_{k_x} \Bigl ( (i \omega + \mu) \matrixX{1} - \matrixX{T}(k_x)-\matrixX{\Sigma}(i \omega) \Bigr )^{-1}.
\end{equation*}
It couples the different impurity problems and encodes the reduced coordination on the edge via $\matrixX{T}$, which is the Fourier-transform of the
hopping and hybridization term plus on-site energy. It takes the form
\begin{align*}
\matrixX{T}(k_x) &= \resizebox{!}{0.04\vsize}{$\begin{pmatrix}
H_{0,0} & H_{0,1} & 0 & \cdots\\
H_{1,0} & H_{1,1} & H_{1,2} & \cdots\\
0 & H_{2,1} & H_{2,2} & \cdots\\
\vdots & \vdots & \vdots & \ddots
\end{pmatrix}$}
\end{align*}
where
\begin{widetext}
\begin{equation*}
H_{j,j} = 
\begin{pmatrix}
-2 t_d\, cos(k_x) & 0 & 0 & 2 V^* sin(k_x) \\
0 & -2 t_d\, cos(k_x) & 2 V^* sin(k_x) & 0 \\
0 & 2 V sin(k_x) & \eps_f -2 t_f\, cos(k_x) & 0 \\
2 V sin(k_x) & 0 & 0 & \eps_f -2 t_f\, cos(k_x)
\end{pmatrix},  
H_{j+1,j}=H_{j,j+1}^{\dagger} = 
\begin{pmatrix}
 -t_d & 0 & 0 & i V \\
0 & -t_d & - i V & 0 \\
0 & i V & -t_f &0 & \\
-i V & 0 & 0 & -t_f
\end{pmatrix}.
\end{equation*}
\end{widetext}
The self-energy is diagonal and obtained from
the impurity calculations via the Dyson equation 
\begin{equation*}
\Sigma_{j k}(i \omega) = \delta_{j k} \Sigma_j (i \omega) = \delta_{j k} \Bigl ( (\mathcal{G}_j^{(0)}(i \omega))^{-1} - (\mathcal{G}_j^{}(i \omega))^{-1} \Bigr ).
\end{equation*}

To solve the impurity problems \eqref{action}, we employ the hybridization-expansion formulation of the continuous-time QMC algorithm \cite{PWernerPRL,PWernerPRB}.
This corresponds to an expansion of the partition function in powers of the hybridization around the local, atomic limit:
\begin{align}
\label{eq_Z}
\frac{Z}{Z_{loc}} = \sum \limits_{k = 0}^{+\infty} &\sum \limits_{\{ \sigma, \sigma' \}} \frac{1}{k!}\,
\int \limits_0^{\beta} d\tau_1 \cdots d\tau_k \int \limits_0^{\beta} d\tau'_1 \cdots d\tau'_k \times\\
&\times \bigl < T_{\tau}\, \creOp{f}{\sigma_1}(\tau_1) \annOp{f}{\sigma'_1}(\tau'_1) \cdots
\creOp{f}{\sigma_k}(\tau_k) \annOp{f}{\sigma'_k}(\tau'_k) \bigr >_{loc} \times \notag\\
& \times \frac{1}{k!}\op{det} \left ( \Delta(\tau_i-\tau'_j) \right ) \notag.
\end{align}
The expectation value of the time-ordered product of operators $\bigl < \dots \bigr >_{loc}$ is taken with respect to the local impurity Hamiltonian $\hamilton_{imp}$.\\
The integral of \eqref{eq_Z} is evaluated by constructing a Markov chain of configurations $C$ of operators
$C_ = (\creOp{f}{\sigma_1}(\tau_1), \annOp{f}{\sigma_1}(\tau'_1),\cdots,\creOp{f}{\sigma_k}(\tau_k), \annOp{f}{\sigma_k}(\tau'_k))$,
which is updated by adding or removing pairs of creation and annihilation operators.
To increase the acceptance rate, in particular at low temperatures, we introduce a third kind of update, namely a shift update. To be precise, we randomly pick a single operator from the whole sequence,
which is then displaced along the imaginary time axis by a random shift $\Delta \tau$, i.e. $f_{\sigma_l}(\tau_l) \to f_{\sigma_l}(\tau_l + \Delta \tau)$.\\
The central quantity for the DMFT is the single-particle Green's function, or equivalently the self-energy. The conventional measurement of $\mathcal{G}(\tau)$ in the hybridization expansion CTQMC
consists of accumulating
\begin{equation*}
\mathcal{G}(\tau-\tau') = \sum \limits_{C} \sum_{i,j=1}^{k}\delta(\tau-\tau_i) \delta(\tau'-\tau'_j) \delta_{\sigma_i \sigma_j} M^{(C)}_{ij}
\end{equation*}
where for a given configuration $C$, $M^{(C)}$ is the inverse of the matrix of hybridization functions: $(M^{(C)})^{-1}_{ij} = \Delta(\tau_i-\tau'_j)$.
Finally, the self-energy is obtained via the Dyson equation. However, a more accurate estimate of the self-energy can be
obtained be a different approach based on higher-order correlation function \cite{improved_sigma_Hafermann}. In this method,
and in the case of a local Hubbard interaction $U$, in addition to $\mathcal{G}$, one accumulates the correlation function
\begin{equation*}
\Gamma(\tau-\tau') = \sum \limits_{C} \sum_{i,j=1}^{k}\delta(\tau-\tau_i) \delta(\tau'-\tau'_j) \delta_{\sigma_i \sigma_j}n_{\bar{\sigma}_i}(\tau_i) M_{ij}
\end{equation*}
where $n_{\bar{\sigma_i}}(\tau_i)$ is the instantaneous occupation of the orbital with spin opposite to $\sigma_i$ at time $\tau_i$.
For the local Hubbard $U$, the self-energy is related to this quantity and the impurity Green's function by
\begin{equation*}
\Sigma(i \omega) = U\, \Gamma(i \omega) \mathcal{G}^{-1}(i \omega).
\end{equation*}
An implementation of the hybridization expansion CTQMC solver, including the improved estimator for the self-energy, can be found in
the ALPS package \cite{ALPS2.0,ALPS_CTQMC}.

The analytical continuation of the results from Matsubara frequencies to real frequencies is done by
the stochastic analytical continuation algorithm described in Ref. \onlinecite{Beach04a}.   The continuation to the real axis is
done on the level of the self-energy, following  Ref. \onlinecite{Goth_magnetic_impurities_KM}.
First, one subtracts the asymptotic   value of the  self-energy 
\begin{align*}
\Sigma(i \omega) &= \Sigma_0 + \Sigma_1 \frac{1}{i \omega} + \dots\\
\Sigma'(i \omega) &= \frac{\Sigma(i \omega)-\Sigma_0}{\Sigma_1} \sim \frac{1}{i \omega}
\end{align*}
where
\begin{align*}
 \Sigma_0 &= \frac{U}{2} \bigl < n_f \bigr >, \; \Sigma_1 = \frac{U^2}{4} \bigl < n_f \bigr > \bigl (2- \bigl < n_f \bigr > \bigr)
\end{align*}
such that 
\begin{align*}
\Sigma'(z)  =  -\frac{1}{\pi} \int  {\rm d} \omega \frac{ \op{Im} \Sigma'(\omega^{+}) }{z - \omega }.
\end{align*}
Analytical continuation provides $ \op{Im} \bigl [ \Sigma'(\omega^{+}) \bigr ]$  from the knowledge of the Matsubara frequency QMC data, and from which one can readily compute 
$\Sigma(\omega^{+})$ on the real frequency axis as well as the Green function:
finally the momentum-resolved Green's function $\matrixX{G}(k_x,\omega^{+})$ and orbital-resolved spectral function $A_j(k_x,\omega)$ on the real axis are obtained by
\begin{align*}
\matrixX{G}(k_x,\omega^{+}) &= \Bigl ( (\omega^{+}  + \mu) \matrixX{1} - \matrixX{T}(k_x)-\matrixX{\Sigma}(\omega^{+}) \Bigr )^{-1}\\
A_j(k_x,\omega) &= -\pi^{-1} \op{Im}\, \bigl [ \matrixX{G}(k_x,\omega^{+}) \bigr ]_{jj}\,.
\end{align*}

\begin{acknowledgments}
 We would like to thank C.-H. Min  as well as F.  Reinert for discussion. 
Funding from the DFG under the grant number  AS120/6-2 (Forschergruppe FOR 1162) is acknowledged. 
We thank the  J\"ulich Supercomputing Centre  and the Leibniz-Rechenzentrum in Munich for generous allocation of CPU time.

\end{acknowledgments}

\bibliographystyle{prsty}
\bibliography{journal_short,tki,fassaad}{}

\end{document}